\documentclass[aps,prb,twocolumn,showpacs,superscriptaddress]{revtex4}
\usepackage{amssymb,amsmath}
\usepackage{graphicx}
\usepackage{multirow}
\usepackage{hyperref}
\usepackage{appendix}
\usepackage{mathtools}
\usepackage{amsmath}
\usepackage{siunitx}
\graphicspath{{./}}

\begin{document}

\title{Ab initio Study of Luminescence in Ce-doped Lu$_2$SiO$_5$: The Role of Oxygen Vacancies on Emission Color and Thermal Quenching Behavior}

\author{Yongchao Jia}
\email[]{yongchao.jia@uclouvain.be}
\affiliation{European Theoretical Spectroscopy Facility, Institute of Condensed Matter and Nanosciences, Universit\'{e} catholique de Louvain, Chemin des \'{e}toiles 8, bte L07.03.01, B-1348 Louvain-la-Neuve, Belgium}
\author{Anna  Miglio}
\affiliation{European Theoretical Spectroscopy Facility, Institute of Condensed Matter and Nanosciences, Universit\'{e} catholique de Louvain, Chemin des \'{e}toiles 8, bte L07.03.01, B-1348 Louvain-la-Neuve, Belgium}
\author{Masayoshi Mikami}
\affiliation{Functional Materials Design Laboratory, Yokohama R$\&$D Center, 1000,
Kamoshida-cho Aoba-ku, Yokohama, 227-8502, Japan}
\author{Xavier Gonze}
\affiliation{European Theoretical Spectroscopy Facility, Institute of Condensed Matter and Nanosciences, Universit\'{e} catholique de Louvain, Chemin des \'{e}toiles 8, bte L07.03.01, B-1348 Louvain-la-Neuve, Belgium}

\date{\today}

\begin{abstract}
We study from first principles the luminescence of Lu$_2$SiO$_5$:Ce$^{3+}$ (LSO:Ce), a scintillator widely used in medical imaging applications, and establish the crucial role of oxygen vacancies (V$_O$) in the generated spectrum. The excitation energy, emission energy and Stokes shift of its luminescent centers are simulated through a constrained density-functional theory method coupled with a ${\Delta}$SCF analysis of total energies, and compared with experimental spectra. We show that the high-energy emission band comes from a single Ce-based luminescent center, while the large experimental spread of the low-energy emission band originates from a whole set of different Ce-V$_O$ complexes together with the other Ce-based luminescent center.  Further, the luminescence thermal quenching behavior is analyzed. The $4f-5d$ crossover mechanism is found to be very unlikely, with a large crossing energy barrier (E$_{fd}$) in the one-dimensional model. The alternative mechanism usually considered, namely the electron auto-ionization, is also shown to be unlikely. In this respect, we introduce a new methodology in which the time-consuming accurate computation of the band gap for such models is bypassed. We emphasize the usually overlooked role of the differing geometry relaxation in the excited neutral electronic state Ce$^{3+,*}$ and in the ionized electronic state Ce$^{4+}$. The results indicate that such electron auto-ionization cannot explain the thermal stability difference between the high- and low-energy emission bands.   Finally, a hole auto-ionization process is proposed as a plausible alternative. 
With the already well-established excited state characterization methodology, the approach to color center identification and thermal quenching analysis proposed here can be applied to other luminescent materials in the presence of intrinsic defects.
\end{abstract}


\maketitle

\section{Introduction}
\label{intro}
During the past two decades, Ce$^{3+}$-doped Lu$_2$SiO$_5$ (LSO:Ce) has attracted a lot of attention from academy and industry.  It has superior characteristics for scintillation, such as a high density, short decay time, high light output,  satisfactory energy resolution and is mechanically and chemically stable. \cite{1992,1993,2001,2003,2005-Kolk,2009,2014}  These properties make LSO, as well as the closely related
Ce$^{3+}$-doped Lu$_{1.8}$Y$_{0.2}$SiO$_5$ (LYSO:Ce),
 leading commercial scintillators for applications in positron emission tomography  and medical imaging equipment. 

LSO crystallizes in a monoclinic crystal structure, with space group of C2/c.  There are two inequivalent Lu crystallographic sites in its crystal structure, which are coordinated by seven and six oxygen atoms, and denoted as Lu1 and Lu2, respectively. \cite{pidol2006,Jia-2018} Due to the same formal valence charge and similar ionic radii of Ce and Lu,\cite{shannon} the dopant Ce ions are expected to occupy the Lu sites. The $4f-5d$ optical transitions of the dopant Ce$^{3+}$ ions dominate the optical spectra of LSO:Ce. 

In a pioneering study,  Suzuki and coworkers\cite{1993} measured the photoluminescence spectra
of LSO:Ce at temperatures from 11~K to 400~K. Their results indicate the presence of two distinct luminescent characteristics in LSO:Ce, with different emission colors and thermal quenching behaviors. The higher energy emission band peaking at around 390~nm shows a typical characteristic of single Ce$^{3+}$ center (at 11~K), namely, the doublet bands of $^2$F$_{7/2}$ and $^2$F$_{5/2}$ ground state levels with an energy  difference of about 2026~cm$^{-1}$, which is agreement with the theoretical value, about 2000~cm$^{-1}$. In addition, the high-energy emission band has a relative good thermal stability, as its thermal quenching temperature, $T_{0.5}$,\cite{Dorenbos2005} is above 300~K. On the other hand, the lower energy emission band exhibits quite different optical performances. First, the low-energy emission band is very broad around its peak (480~nm ), without the typical doublet characteristic of the single Ce$^{3+}$ center, even at very low temperature, 11~K. 
Second, this broad emission band suffers from a very strong thermal quenching, and could not be observed  above 80~K. Suzuki \textit{et al} proposed a two-activation-center model to explain these two luminescence characteristics, each originating from a
different set of Ce$^{3+}$ centers (referred to as Ce1 and Ce2) occupying the crystallographically independent Lu1 and Lu2 sites, respectively.  However, such model failed to explain the absence of doublet structures in the low-energy emission band.

\begin{figure*}
\includegraphics[scale=0.125]{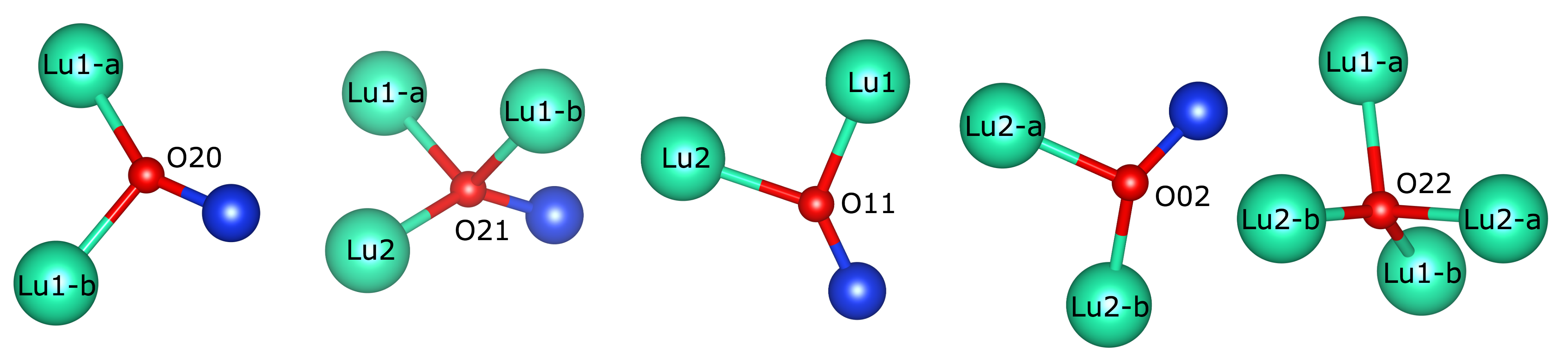}
\caption{LSO:Ce defect sites. The Ce atom can substitute Lu at each of the sites shown in green, while the neighboring oxygen atom, in red, can be removed, giving thirteen combinations of Ce-V$_O$. The Si atoms are shown in blue.}\label{figure_cry}
\end{figure*}

A few years later, a different assignment of the two luminescent centers was proposed by Naud \textit{et al},\cite{1996-a} the high- and low- energy emission bands corresponding to Ce$^{3+}$ ions at the substitutional and interstitial Lu sites, respectively. This preliminary assignment assumed that Ce$^{3+}$ ions at two substitutional Lu sites give very similar optical behavior, while the absence of doublet structures in the low-energy emission band is due to the variation of the interstitial Ce$^{3+}$ local environments. Until now, there is no direct evidence available for these assumptions. Furthermore, the assignment of the low-energy emission band to the interstitial Ce$^{3+}$ ions was not shown to yield a red-shift of emission spectra and strong thermal quenching compared to the high-energy emission peak. But more fundamentally, such interstitial Ce$^{3+}$ ions should be very difficultly to form, due to the large size of Ce$^{3+}$ ion and lack of large cavities in LSO crystal structure.

Recent achievements in the study of LSO:Ce may provide new clues on the origin of the two emission bands.\cite{2014,TL-1994,1998-1,ab-3,ab-2,m2-1,co-dpoing-1} These studies indicate that intrinsic oxygen vacancies are highly probably formed in the synthesis process of LSO:Ce under oxygen-poor atmosphere (Ar). Their existence was proposed to be the physical origin of the afterglow and thermoluminescence (TL) phenomenon in LSO:Ce,\cite{2014,TL-1994,1998-1,ab-3,ab-2} as well as the effect of co-doping with divalent cation ion (M$^{2+}$) on the scintillating performance of LSO:Ce.\cite{m2-1,co-dpoing-1} These facts suggest that oxygen vacancies can be closely linked to the luminescence in LSO:Ce and should be reasonably considered for the assignment of the two emission bands, instead of the interstitial Ce$^{3+}$ ions. 

In our previous work,\cite{Jia-2018} we have studied from first principles the stability of neutral and charged oxygen vacancies in LSO. We found that in LSO oxygen vacancies are most likely to be formed as neutral vacancies within a [SiO$_4$] tetrahedron instead of interstitial oxygen sites bonded to Lu atoms only. Similar results were obtained for the  isostructural compound Y$_2$SiO$_5$.
In addition, the incorporation of Ce$^{3+}$ ion in LSO was shown to have a negligible effect on the stability of oxygen vacancies.

In the present work we analyze the Ce$^{3+}$ luminescence in LSO from first principles, which includes estimating the effect of intrinsic defects (oxygen vacancies) on the luminescence, especially on the transition energies and thermal stability. 
We assign to specific luminescent centers the two emission bands in LSO:Ce. While the luminescent center for the high-energy emission band is simply due to substitutional Ce$^{3+}$ ions at Lu1 site, the very broad low-energy emission band originates from the contribution of ten Ce$^{3+}$ luminescent centers, one being substitutional Ce$^{3+}$ ions at Lu2 site and nine being different combinations of one substitutional Ce$^{3+}$ ion and one neutral oxygen vacancy. The thermal quenching behavior of the low-energy band is analyzed and the results indicate that the two usual explanatory mechanisms, $4f-5d$ crossover and auto-ionization, cannot easily account for its low thermal stability within our methodology. We put forward the hole auto-ionization process as a possible candidate for the strong thermal quenching of the low-energy emission band. 

The basic methodology to study excitation energy, emission energy and Stokes shift, 
is based on the constrained DFT method (CDFT) and $\Delta$SCF, already used e.g. in the works of Canning and coworkers,\cite{canning2011,canning2014} and our previous studies of luminescence\cite{Jia-2016,Ce,Jia-2017}.
The further methodology to study thermal quenching via $4f-5d$ crossover was considered in a first-principle
context in our previous work.\cite{Ce,Jia-2017} In addition to these first principles methodologies, 
our present study of the thermal quenching due to 
electronic auto-ionization, relies first on a procedure to bypass the difficult calculation of the band gap,
focusing instead on the \textit{relative differences} of activation energy, and second, on the quantification of the effect of geometry relaxation
from the neutral electronic excited state to the ionized state, overlooked in previous studies, including ours.\cite{ponce-2016,Jia-2017}

Furthermore, we explore the hole auto-ionization model from first principles, which enlarges the global picture of thermal quenching mechanisms of luminescent materials. 
Indeed, the previous works about thermal quenching in rare-earth doped phosphors only focused on the $4f-5d$ crossover and electron auto-ionization models, 
while the hole auto-ionization model was only very recently proposed by Dorenbos\cite{dorenbos5} and qualitatively analyzed by him. The methodology and mechanism proposed here are not specific to LSO:Ce, but can be reasonably applied to the analysis of structure-property relationship in other luminescent materials.

As a whole, this paper provides a basic framework for the study of intrinsic defect effects on Ce$^{3+}$ luminescence that can be applied to any host material. This topic indeed is largely ignored in the previous research, from both theory and experiment. The method used to analyze the auto-ionization model can be easily applied to the color center identification in multi-site compounds, which occurs quite often in the area of scintillators, LED phosphors and solid-state lasers.

This paper is structured as follows. In Section~\ref{num_approach}, we briefly describe the theoretical methods used in the calculation of the Ce$^{3+}$ luminescence characteristics. Then, the results for Ce$^{3+}$ luminescence at the standard Lu sites are shown in Section~\ref{Ce}. The effect of oxygen vacancies on Ce$^{3+}$ luminescence is depicted in Section~\ref{Ce-Vo}. The thermal quenching of Ce$^{3+}$ luminescence in LSO is analyzed in Section~\ref{thermal}. Finally, the conclusions are presented in Section~\ref{conclusion}.

\section{Numerical approach}
\label{num_approach}

Calculations were performed within density functional theory (DFT) using the projector augmented wave (PAW) method as implemented in the ABINIT package.\cite{Blochl1994,Abinit2009,Gonze2016,Marc2008} Exchange-correlation (XC) effects were treated within the generalized gradient approximation (GGA).\cite{Perdew1996}  For the Ce$^{3+}$ doped calculations, DFT(PBE)+U was used, allowing the Ce$_{4f}$ states to be found inside the band gap. From our previous study about Ce-doped phosphors,\cite{Jia-2016,Ce} we found that by increasing the  U value from 4 eV to 5 eV the change on the calculated transition and relaxation energies is within 0.05 eV. In the present work, we set  U = 4.6 eV  and J = 0.5 eV as in our previous study.

The PBE atomic datasets involved in present paper are the same as in our previous work on LSO.\cite{Jia-2018} 

For structural relaxation and band structure calculations, the convergence criteria were set to 10$^{-5}$ Ha/Bohr (for residual forces) and 0.5 mHa/atom (for the tolerance on the total energy), corresponding to a kinetic energy cutoff of 30~Ha and a 2$\times$4$\times$4 Monkhorst-Pack sampling of Brillouin Zone. 

Detailed convergence studies were conducted on the supercell size used for the defect calculations. The data presented in the following were obtained using a 64-atom supercell, which was found to be sufficient to converge defect formation energies to better than 0.05~eV, compared to those obtained on a 128-atom supercell. 
The calculations involving the Ce$^{3+}$ ion were performed with one Ce atom substituting one Lu atom both in the bulk supercell (Lu$_{16}$Si$_{8}$O$_{40}$)  and in the 64-atoms supercell containing one oxygen vacancy (Lu$_{16}$Si$_{8}$O$_{39}$). In the present work, we focus on the situation in which the oxygen vacancy is formed nearby the Ce$^{3+}$ ions, to evaluate the effect of oxygen vacancy on Ce$^{3+}$ luminescence. Thus, a total of thirteen combinations of Ce-V$_O$ (7+6 coordinated oxygen sites nearby two Ce$^{3+}$) are considered (see Fig.~\ref{figure_cry}).

\begin{figure}
\includegraphics[scale=0.35]{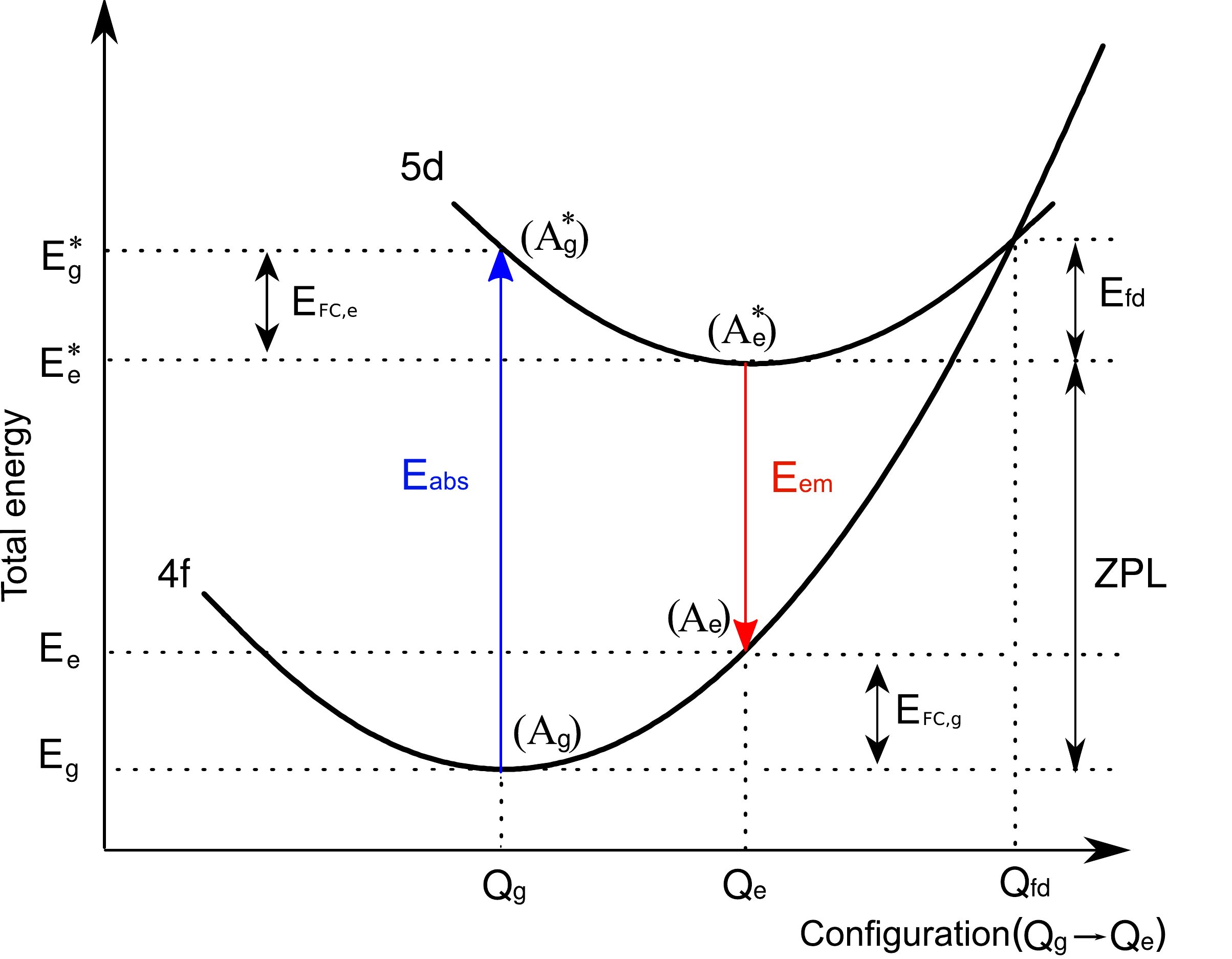}
\caption{The one-dimensional configuration coordinate diagram.}\label{figure_ccd}
\end{figure}

The calculation of the $4f-5d$ neutral excitation of the Ce$^{3+}$ ion is based on the constrained DFT method (CDFT), following the works of Canning and coworkers,\cite{canning2011,canning2014} and our previous studies of luminescence\cite{Jia-2016,Ce,Jia-2017}. The electron-hole interaction, an essential contribution for the study of neutral excitations, is mimicked by promoting the Ce$_{4f}$ electron to the Ce$_{5d}$ state, constraining the fourteen $4f$ bands to be unoccupied, while occupying the lowest of the $5d$ state. Following the CDFT method, the absorption and emission energies of LSO:Ce are calculated and compared with experiment. 
The energy barrier of thermal quenching via $4f-5d$ crossover, E$_{fd}$, and the criterion for immediate non-radiative recombination,  $\Lambda$, are analyzed following the framework of a one-dimensional configuration coordinate diagram (1D-CCD),\cite{book1,book2} together with the information obtained from CDFT. The notation related to the 1D-CCD  is illustrated in Fig.~\ref{figure_ccd} and detailed information on the method implementation and further comments can be found in our previous works.\cite{Ce,Jia-2017}

\begin{figure}
\includegraphics[scale=0.35]{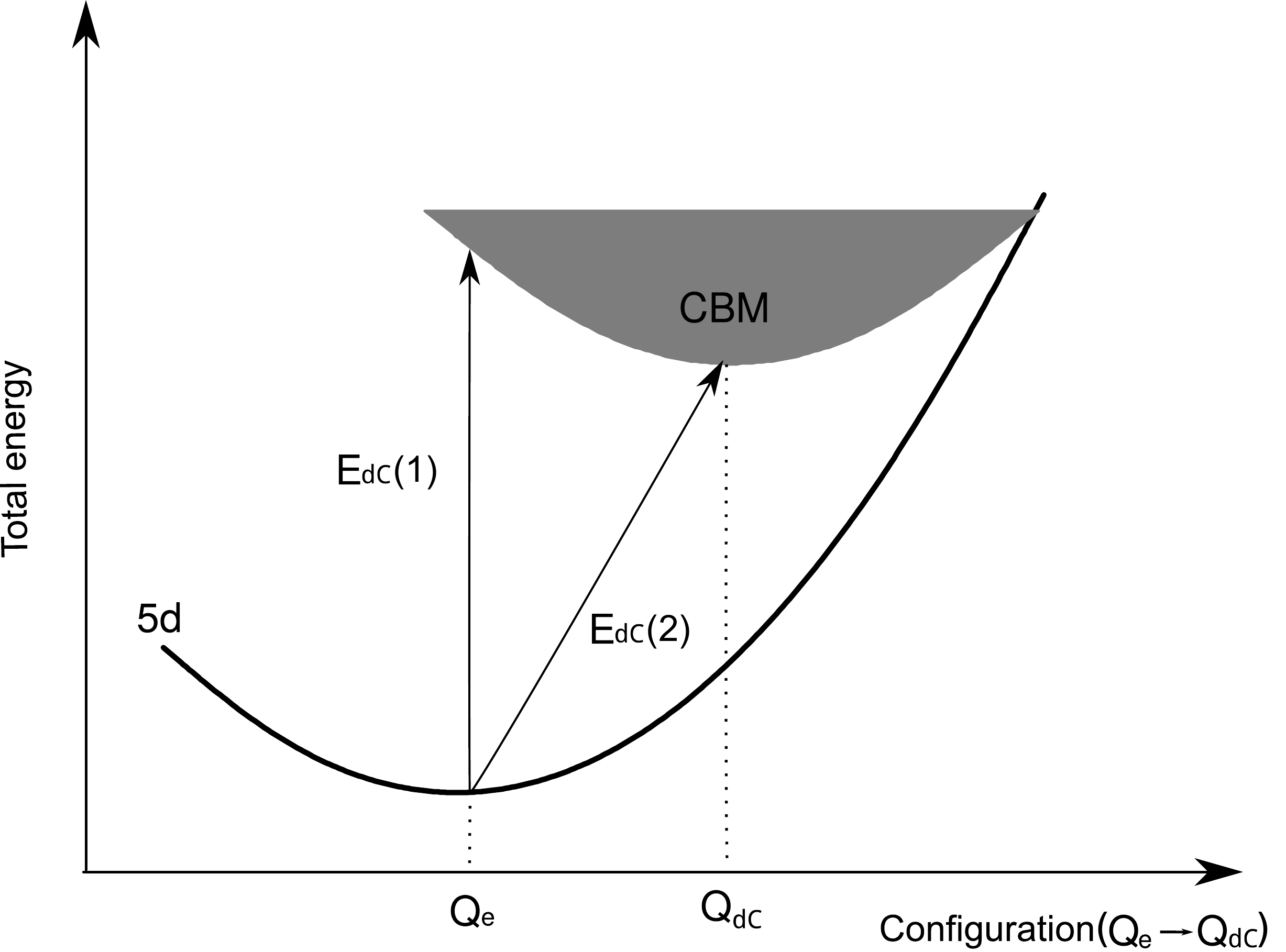}
\caption{Configuration coordinate diagram for the auto-ionization model. E$_{dC}$(1) is the (optical) energy difference corresponding to transition from the Ce$_{5d}$ excited state at its minimum energy geometry (Q$_e$)
to the delocalized state - without geometry relaxation, while E$_{dC}$(2) is the (thermal) energy difference in which the final state includes geometry relaxation to its minimum energy geometry Q$_{dC}$.}\label{figure_auto}
\end{figure}

Another proposed mechanism for thermal quenching is described by the auto-ionization model,\cite{huang1950,Dorenbos2005,ponce-2016,Ueda2017}  sketched in Fig.~\ref{figure_auto}. This model relies on the possibility for the electron in the excited state to be driven into the conduction band, e.g. by thermal lattice vibrations, resulting in its delocalization, after which the radiative recombination rate drops considerably, and other non-radiative recombination mechanisms have the time to occur. The electron delocalization usually results in a geometry relaxation in which the local collective displacement of atoms from Q$_e$ to Q$_{dC}$ might be quite different from the one observed when going from Q$_g$ to Q$_e$.
The total energy of both the Ce$_{5d}$ and the delocalized electron states changes with the collective displacement of atoms from Q$_e$ to Q$_{dC}$. It reaches its minimum at  Q$_e$ when the electron occupies the Ce$_{5d}$ state, and at Q$_{dC}$ when the electron is delocalized. 
Consequently, in Fig.~\ref{figure_auto} we distinguish the configuration coordinate between Q$_e$ $\rightarrow$ Q$_{dC}$ from the one of Q$_g$ $\rightarrow$ Q$_e$, at variance with our earlier works, Fig. 1 of Ref.[\onlinecite{ponce-2016}] and Fig. 2 of Ref.[\onlinecite{Jia-2017}]. Also, the CBM zone, in grey in Fig.~\ref{figure_auto}, is delimited by a parabola in configuration coordinate diagrams because they represent total energies and not electronic energies, and one can thus distinguish an optical (fixed geometry, Q$_e$) and a thermal (relaxed geometry, Q$_e$ $\rightarrow$ Q$_{dC}$) energy difference between the total energy of the Ce$_{5d}$ electron situation and the total energy of the delocalized CBM electron situation. The need to go beyond the 1D-CCD model has already been pointed out in the $4f-5d$ crossover context, \cite{Bartram1986-1,Bartram1986-2} as will be discussed later.

Some previous publications considered the auto-ionization mechanism in LSO:Ce in conjunction with experimental results.\cite{2001,2005-Kolk} In their analysis of experimental data, those studies did not mention the Q$_e$ $\rightarrow$ Q$_{dC}$ geometry relaxation, making the physical picture of thermal quenching barrier for the auto-ionization model incomplete. 
Also, an incorrect flat representation of the CBM zone in configuration coordinate diagrams is often used, see e.g. the Fig. 4 in Ref.[\onlinecite{2005-Kolk}].

From the theoretical viewpoint, the calculation of the absolute energy barrier for the thermal quenching via auto-ionization model,  E$_{dC}$, is not an easy task. Indeed, because it depends on the band gap, its accurate estimation would need to go beyond DFT-PBE, which is subject to the well-known DFT band gap problem, e.g. using hybrid functionals or the GW approximation. Also, the effect of geometry relaxation (from Q$_e$ to Q$_{dC}$) on the absolute energy barrier of thermal quenching should be taken into account. 

In the present study, we try to understand the effect of oxygen vacancies on thermal stability of different Ce$^{3+}$ ions associated with oxygen vacancies, in their electronic excited state. Instead of focusing on the absolute value for the energy barrier within the auto-ionization model,  we will actually compare variations of E$_{dC}$ between the different luminescent centers,
based on the optical (fixed geometry) and thermal (relaxed geometry) transition energy of $\varepsilon$($Ce^{3+,*}/Ce^{4+}$), indeed taking geometry relaxation effects into account. Thus we bypass the
band-gap problem, but include the geometry dependence, as follows.

E$_{dC}$ can be calculated from the difference between the conduction band energy  and the so-called transition energy level $\varepsilon$($Ce^{3+,*}/Ce^{4+}$):\cite{Van2014}
\begin{equation}
\label{E_dC}
E_{dC} =  \varepsilon(c) - \varepsilon(Ce^{3+,*}/Ce^{4+}), 
\end{equation}
The transition energy level is the Fermi level,  $\epsilon_{f}$ when the formation energies of Ce$^{3+,*}$ and Ce$^{4+}$ ions are equal. They are obtained from
 \begin{equation}
\label{formation_e3}
E_\text{f}(Ce^{3+,*}) = E_{tot}(Ce^{3+,*}) - E_{tot}(bulk) +  \mu_{Lu} -  \mu_{Ce},
\end{equation} 
and
 \begin{eqnarray}
\label{formation_e4}
E_\text{f}(Ce^{4+}) &=& E_{tot}(Ce^{4+}) - E_{tot}(bulk) \nonumber \\
                              &+&  \mu_{Lu} -  \mu_{Ce} + \epsilon_{f} + E_{corr},
\end{eqnarray} 
where E$_{tot}$(Ce$^{3+,*}$) is the total energy of the (neutral) supercell with excited state Ce$^{3+,*}$ ion, E$_{tot}$(Ce$^{4+}$) is the total energy of the (charged) supercell with Ce$^{4+}$ ion,
E$_{tot}$(bulk) is the total energy of the undoped LSO supercell. E$_{corr}$ gathers the
correction terms that are needed when charged systems are treated with periodic supercells, which is the case for
the Ce$^{4+}$ ion supercell.
In the present work, we consider the charged monopole correction, which is the dominant contribution,
described by a Madelung-type formula.\cite{Chris-2009} The calculation details can be found in our previous work.\cite{Jia-2018}

Computing the transition energy level by equating Eq.[\ref{formation_e3}] and Eq.[\ref{formation_e4}] yields:\begin{equation}
\varepsilon(Ce^{3+,*}/Ce^{4+})  = E_{tot}(Ce^{3+,*}) - E_{tot}(Ce^{4+}) - E_{corr}. 
\end{equation}

For E$_{tot}$(Ce$^{4+}$), one can work with the geometry from the excited state Ce$^{3+,*}$ system (Q$_e$), 
giving the optical E$_{dC}$(1) or from the relaxed geometry of the ionized state Ce$^{4+}$ system ($Q_{dC}$)
giving the thermal E$_{dC}$(2). Adding a delocalized electron to the latter does not change the relaxed geometry.
The formulation Eq.[\ref{E_dC}] allows us to bypass the DFT band-gap problem, in that we will compare $E_{dC}$ values for different defect states, for which the contribution
$\varepsilon(c)$ to Eq.[\ref{E_dC}] is the same, while the remaining $\varepsilon(Ce^{3+,*}/Ce^{4+})$
contribution is not affected by the energy of the conduction band with respect to the valence band, 
or with respect to the Fermi energy.

Then, the differences between E$_{dC}$(1) and E$_{dC}$(2) of Ce$^{3+}$ luminescence, with and without the presence of an oxygen vacancy surrounding, can be calculated via Eq.[\ref{E_dC}], be
it for the optical or thermal case, through the computation of the transition transition energy level, $\varepsilon$($Ce^{3+,*}/Ce^{4+}$, Q$_e$) and $\varepsilon$($Ce^{3+,*}/Ce^{4+}$, Q$_{dC}$), respectively.

\section{Results and Discussions}
\label{result}
\subsection{Ce$^{3+}$ luminescence at substitutional Lu sites}
\label{Ce}

Before considering the luminescence, let us recall the results obtained for the formation energy
of substitutional Ce$^{3+}$ ions at the two Lu$^{3+}$ sites of LSO,
from our previous work.\cite{Jia-2018} It was found that the Ce$^{3+}$ ion at Lu1 site has a lower energy (45.93~kJ/mol=0.47eV) than when it substitutes
at Lu2 site (see the row related to A$_g$ in Table~\ref{table-1}). 
This preference for Ce$^{3+}$ occupation is correlated to the larger volume of the Lu1 coordination polyhedron than the one of the Lu2 site. In LSO, the Lu-O distances of Lu2 site range from 2.16 to 2.24 \AA. The other Lu1 site exhibits a distribution of six Lu-O distances between 2.16 and 2.34 \AA, 
while the seventh oxygen is at 2.61\AA, thus it can be considered as a `6+1' oxygen environment with a larger space. The ionic radius of Ce$^{3+}$ ion is slightly larger than the one of Lu$^{3+}$ ion. As a result, Ce$^{3+}$ ions can be expected to enter the Lu$^{3+}$ site that is the least compact one. Our assessment is consistent with previous experimental and theoretical studies.\cite{2006-0,1997-0} 
In addition, we found that the Ce$^{3+}$-doped DFT+U calculations
describe the Ce$_{4f}$ electronic state inside the band gap, which is the prerequisite for the scintillation. 

\begin{figure*}
\includegraphics[scale=0.4]{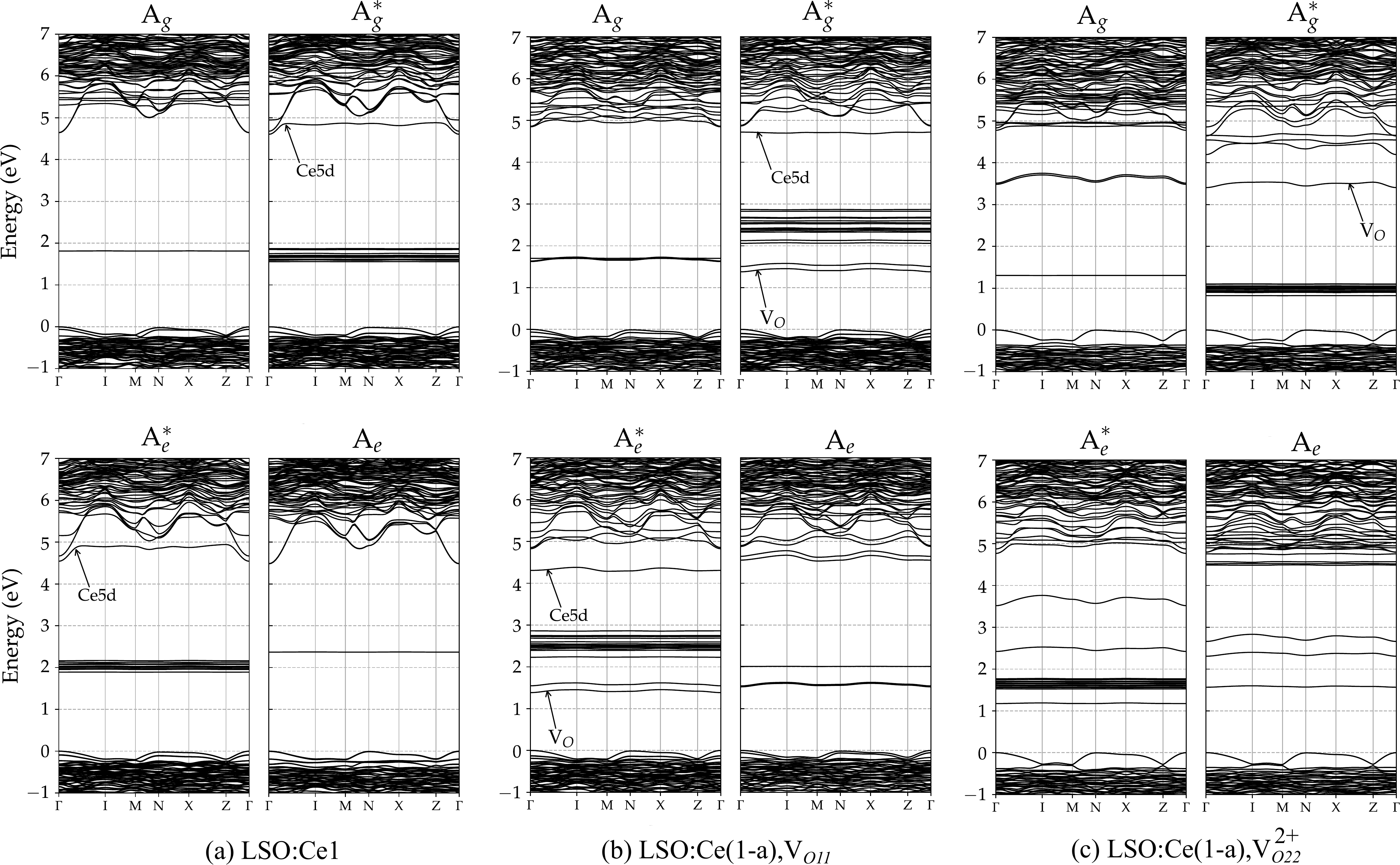}
\caption{Electronic band structures of LSO:Ce,V$_O$  in the four cases (A$_g$, A$_g^*$, A$_e^*$ and A$_e$, as labelled above each band structure) corresponding
to ground-state and excited state geometries and ground-state and excited state electronic occupancies, as depicted in Fig.~\ref{figure_ccd}. Three geometries are considered, giving each four band structures.}\label{figure_LSO-Ce}
\end{figure*}

Based on the ground state geometries of Ce$^{3+}$ ions in LSO, the electronic band structures of LSO:Ce  (Ce at Lu1 and Lu2 sites) have been calculated with the CDFT method, then
the atomic positions have been relaxed in the CDFT electronic excited state, and ground and excited state total energies and electronic band structures have been computed in this geometry. 
The band structures for the LSO:Ce1 case is shown in the Fig.~\ref{figure_LSO-Ce}. 
Compared to the result for LSO bulk,\cite{Jia-2018} the incorporation of Ce$^{3+}$ ion induces energy states inside the band gap: one occupied Ce$_{4f}$ state for the ground states (A$_g$ and A$_e$), fourteen unoccupied Ce$_{4f}$ states and one occupied Ce$_{5d}$ state for the excited states (A$_g^*$ and A$_e^*$). The meanings of A$_g$, A$_e$, A$_g^*$ and A$_e^*$ are the same as in our previous work,\cite{Ce,Jia-2017} see Fig.~\ref{figure_ccd}, and the changes of electronic band structures are also similar with the results for Ce$^{3+}$-doped phosphors.\cite{Ce} The situation for LSO:Ce2 is similar as well (not shown).  Thus, the $4f-5d$ neutral excitation of Ce$^{3+}$ ions substitutional Lu sites of LSO is reasonably well described within CDFT. Therefore, the absorption and emission energies and Stokes shift (Table~\ref{table-1}) have been calculated from the total energy differences between the ground and excited states. 

As mentioned in the introduction, Suzuki \textit{et al} characterized the Ce$^{3+}$ luminescence in LSO and observed two emission bands in the PL spectra. The high-energy emission band, peaking at 390~nm (3.18 eV), showed a typical characteristic of single Ce$^{3+}$ center. The lowest excitation energy of this high-energy emission band was observed at 356~nm (3.48~eV). The other low-energy emission band, peaking at 480~nm (2.68~eV) was very broad, without showing the  typical characteristic of single Ce$^{3+}$ center even at very low temperature, 11~K. Its lowest excitation energy had a peak around 376~nm (3.30~eV). 

Taking into account the slight overestimation of the computed absorption and emission energies in our
technique, on the order of 0.2 eV \cite{Jia-2017}, the higher energy emission peak and the corresponding
excitation threshold, can be reasonably assigned to LSO:Ce1, which is also the preferred substitutional site 
for Ce. The small width (so single Ce$^{3+}$ character) is also coherent with the presence of a single luminescent center for this emission peak.

The origin of the low-energy emission band is more subtle. The calculated transition energy 
of LSO:Ce2 agrees with the low-energy emission band from experiment.
So, the Ce2 site might give some contributions to this lower energy emission peak. 
However, the absence of doublet structures in the low-energy emission band indicates that such band cannot just come from a single luminescence center. Also, the Ce$^{3+}$ occupying the Lu2 site 
has a total energy of 45.93~kJ/mol higher than the LSO:Ce1. 
There must be other Ce$^{3+}$ ions in non-standard sites, giving an emission energy in the region of the lower broad peak. 
Even if the low-energy emission band has not been completely determined as of now, 
our calculation results indicates the hypothesis from Naud \textit{et al}\cite{1996-a},
namely, the similar emission energy for LSO:Ce1 and LSO:Ce2,  is not correct: the luminescence of Ce$^{3+}$ from the two substitutional Lu sites is quite different.

\begin{table}[hbtp]
\centering \renewcommand\arraystretch{1.5}
\caption{Calculated total energies, transition energies and Stokes shift of LSO:Ce. The total energies of A$_g$,  A$_g^{*}$, A$_e^{*}$, and A$_e$ are shown as the relative difference with respect to the E$_g$ value of LSO:Ce1, -36878.28~eV.}
\label{table-1} \begin{tabular}{ccc} \hline
           & LSO:Ce1 & LSO:Ce2 \\ \hline
E$_g$ &  0.00 eV & 0.47 eV \\
E$_g^{*}$& 3.80 eV  &  4.09 eV \\
E$_e^{*}$& 3.58eV &  3.77 eV \\
E$_e$&   0.26 eV & 0.92 eV \\\hline
$\Delta$E$_{abs}$(E$_g^{*}$-E$_g$)&  3.80 eV & 3.62 eV \\ 
$\Delta$E$_{em}$(E$_e^{*}$-E$_e$)&  3.32 eV & 2.85 eV \\
$\Delta$S(Cal.)& 3871 cm$^{-1}$ & 6210 cm$^{-1}$ \\\hline
\end{tabular} \end{table}

\subsection{The effect of oxygen vacancy (V$_O$) on Ce$^{3+}$ luminescence}
\label{Ce-Vo}

In the previous section,  the higher energy emission peak has been assigned to the LSO:Ce1 site. In this section, we identify the luminescent centers for the low-energy emission band. We focus on the effect of oxygen vacancies on the Ce$^{3+}$  luminescence. As described in Section~\ref{intro}, the existence of oxygen vacancies in  LSO  was first assumed based on the afterglow and TL behaviors of LSO:Ce scintillator, which might be due to the excited Ce$_{5d}$ electron and/or Ce$_{4f}$ hole being trapped in the oxygen vacancy. In our previous study,\cite{Jia-2018} we have calculated the relative stability of five kinds of oxygen vacancies, with different exchange-correlation functionals and different pseudopotentials, for several supercells and charged states. Let us summarize the conclusion from this study in this respect.\cite{Jia-2018}  
First, the formation energy of V$_O^{+}$ is higher than those of V$_O$ and V$_O^{2+}$, in 
nearly all the range of permitted values for the Fermi level. Therefore, V$_O^{+}$  should not be stable in LSO(:Ce$^{3+}$). Second, for the V$_O$, our previous work indicates that the V$_O$ forms in the [SiO$_4$] tetrahedral site (O$_{20}$, O$_{21}$, O$_{11}$ and O$_{02}$).  The V$_O^{2+}$ prefers to form at interstitial oxygen site bonded to lutetium atoms only (O$_{22}$), and V$_O^{2+}$ at O$_{20}$ give a similar (slightly higher) formation energy with respect to the result of O$_{22}$. As a result, we study the effect of V$_O$  in the [SiO$_4$] tetrahedral site, and V$_O^{2+}$ at O$_{20}$ and O$_{22}$ on the Ce$^{3+}$  luminescence, corresponding to 9 and 6 cases of Ce-V$_O$ combinations, as shown in Fig.~\ref{figure_cry}.

We first evaluate the effect of the neutral oxygen vacancy on Ce$^{3+}$  luminescence. For the nine Ce-V$_O$ combinations, the CDFT method delivered the transition energy for the luminescence of a Ce$^{3+}$
close to a neutral oxygen vacancy. V$_O$ induces a doubly occupied state above the valence band maximum. Fig.~\ref{figure_LSO-Ce}(b) shows a typical set of band structure results, namely for LSO:Ce(1-a),V$_{O11}$.

\begin{figure}
\includegraphics[scale=0.10]{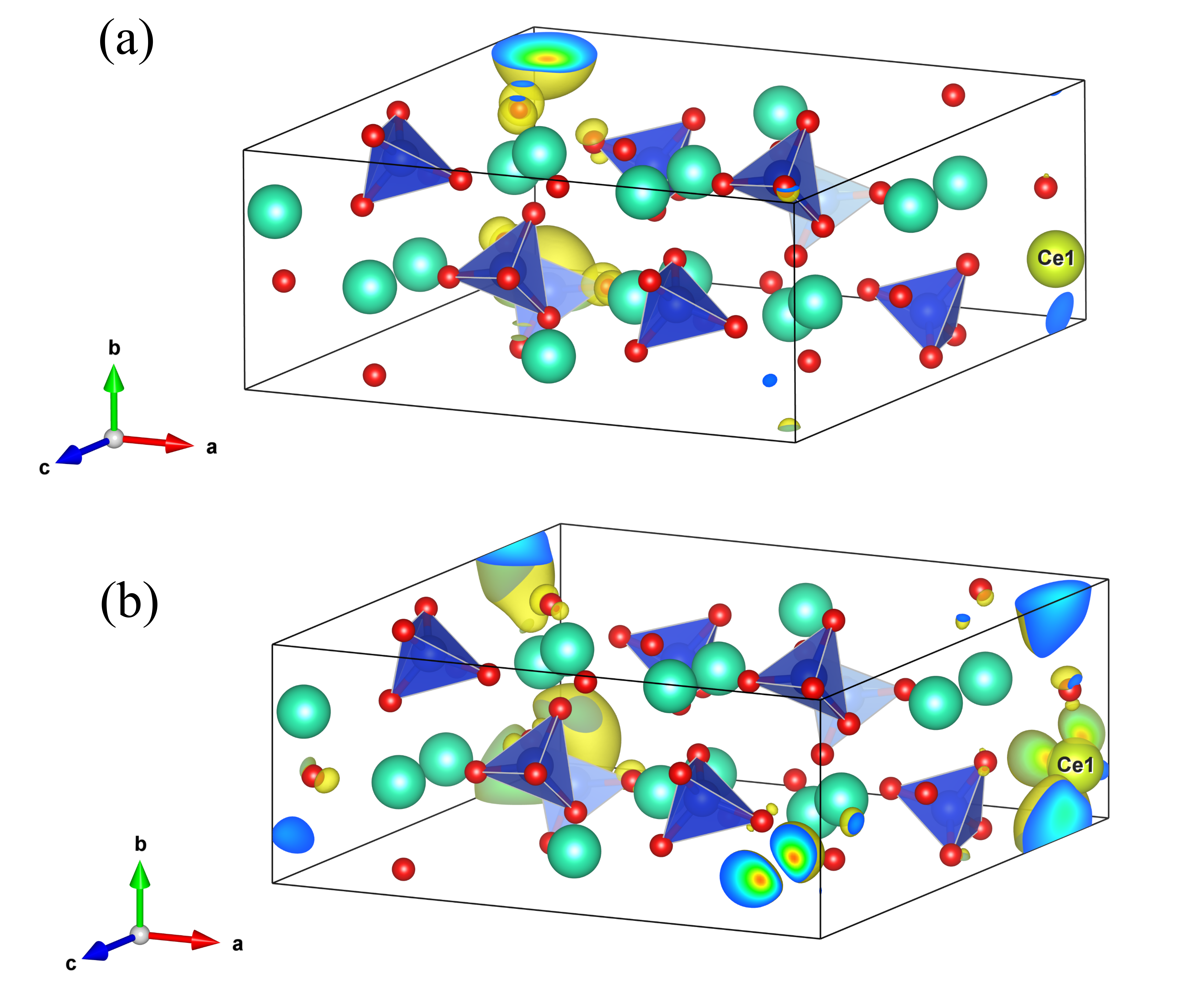}
\caption{Charge density of A$_g^*$ case of LSO:Ce(1-a),V$_{O11}$ at $\Gamma$ point: (a) V$_O$ state; (b) Ce$_{5d}$ state.}\label{figure_charge}
\end{figure}

Such calculation shows that the creation of V$_O$ does not change the nature of the $4f-5d$ neutral excitation of Ce$^{3+}$ ion in LSO. In the band structure for the A$_g^*$  and A$_e^*$ cases, there are fourteen unoccupied Ce$_{4f}$ states and one occupied Ce$_{5d}$ state inside the band gap. Beside these states from Ce$^{3+}$ ions, a doubly occupied state from the oxygen vacancy also occurs in the band gap. Fig.~\ref{figure_charge} shows the charge density of the orbitals corresponding to the
Ce$_{5d}$ and oxygen vacancy states.  The results for the other eight Ce-V$_O$ combinations give a similar change of electronic band structures under excitation (not shown). The transition energies of Ce$^{3+}$  in the nine LSO:Ce,V$_O$  cases are calculated as usual. The band structure results of LSO:Ce,V$_O$ can be compared to those of Fig.~\ref{figure_LSO-Ce}, and will be discussed in the following. 

The results are listed in the Table~\ref{table-2}, together with the experimental data for the lower energy emission peak as well as the calculated values for LSO:Ce1 and LSO:Ce2 for comparison. We first focus on the comparison between the calculated transition energy of LSO:Ce,V$_O$ with the experimental data for the lower energy emission peak. The calculated absorption and emission energies match the experiment data within 0.2~eV, for most cases of LSO:Ce,V$_O$. Indeed, only the absorption energy of LSO:Ce(2-a),V$_{O02}$ gives a larger difference of 0.31~eV. Such small differences justifies our expectation that the low-energy emission band can be related to the intrinsic oxygen vacancy. Also, the average of these nine Ce-V$_O$ results again matches nicely with experiment. Therefore, 
each one of the Ce$^{3+}$ ions at substitutional Lu sites, close to a neutral oxygen vacancy from a [SiO$_4$] tetrahedral site, can contribute to the low-energy emission band. 
The absence of doublet structure from experiment is due to the spread related to multi-center luminescence, as we will discuss later.

\begin{table}[]
\centering\renewcommand\arraystretch{1.5}
\caption{Transition energies (eV) and Stokes shift (cm$^{-1}$) of LSO:Ce,V$_O$. The effect of oxygen vacancy on the Ce$^{3+}$ luminescence in LSO is highlighted as the values of $\Delta$(Abs, V$_O$) and $\Delta$(Em, V$_O$) for absorption and emission energy, respectively. Experimental data for the low-energy emission band is shown for comparison. }
\label{table-2}
\begin{tabular}{lccccc}\hline\hline
                           & Abs & $\Delta$(Abs, V$_O$) & Em & $\Delta$(Em, V$_O$) & Stokes shift \\ \hline
Ce1                     & 3.80      & ---      & 3.32  & ---        & 3871    \\
Ce1-a, V$_{O20}$ & 3.07   & 0.71    & 2.64 & 0.68         & 3468    \\
Ce1-b, V$_{O20}$ & 3.45    & 0.35   & 2.86 & 0.46         & 4759     \\
Ce1-a, V$_{O21}$ & 3.33    & 0.47   & 2.79  & 0.53        & 4372     \\
Ce1-b, V$_{O21}$ & 3.29    & 0.59   & 2.77  & 0.55        & 4194     \\
Ce1, V$_{O11}$   & 3.35      & 0.45   & 2.79  & 0.53        & 4517    \\\hline
Ce2                     & 3.62        & ---    & 2.85   & ---       & 6210     \\
Ce2, V$_{O21}$    & 3.27     & 0.35  & 2.49   & 0.36       & 6291     \\
Ce2, V$_{O11}$    & 3.10     & 0.52   & 2.62  & 0.23        & 3872     \\
Ce2-a, V$_{O02}$ & 2.99     & 0.63   & 2.60  & 0.25        & 3146    \\
Ce2-a,V$_{O02}$ & 3.17      & 0.45   & 2.74  & 0.11        & 3468     \\\hline
Average(V$_O$)               & 3.22      & 0.50     & 2.70  & 0.41        & 4198    \\\hline
Exp.  & 3.30 & --- & 2.68 & --- &  4951 \\ \hline

\end{tabular}
\end{table}

A red-shift of the emission energy when a neutral oxygen vacancy is created close to the Ce1 or Ce2 site
is also observed in Table~\ref{table-2}. Let us discuss the reason for such red-shift. For this task, we examine first the outcome of Dorenbos' semi-empirical model fitted on our first principles structural data.  In the Dorenbos' semi-empirical model, the energy of the first allowed $4f\rightarrow5d$ transition of the  free Ce$^{3+}$ ion is lowered by the crystalline environment (\textit{A}), with a shift denoted \textit{D}(\textit{A}). This lowering is the sum of the spectroscopic red-shift arising from the centroid shift of the Ce$_{\textit{5d}}$ energy, $\varepsilon_{c}$(\textit{A}), and the crystal-field splitting, $\varepsilon_{cfs}$(\textit{A}), of the Ce$_{\textit{5d}}$ states. The calculations of $\varepsilon_{c}$(\textit{A}) and $\varepsilon_{cfs}$(\textit{A}) depend on the electronegativity ($\chi_{av}$) and spectroscopic polarization ($\alpha_{sp}$) of the anion, and on the crystal-field strength parameter ($\beta$) that is related to the shape and size of the anion coordination polyhedron.\cite{dorenbos2,dorenbos3,dorenbos4} A similar analysis for the Ce-doped compounds, using our ab-initio relaxed structures, can be found in our previous work.\cite{Jia-2016}. In Table~\ref{table-3} we list the relevant parameters of Dorenbos' model for LSO:Ce1 and LSO:Ce(1-a),V$_{O20}$. However, this analysis fails to predict the red-shift of the spectra due to the existence of neutral oxygen vacancies. Indeed, Dorenbos' model provides a blue-shift for the transition energy of LSO:Ce(1-a),V$_{O20}$. The creation of oxygen vacancy V$_{O20}$ leads to a blue-shift for the centroid of Ce$_{5d}$ state and the blue-shift is dominated by the increase of the crystal field splitting.

\begin{figure}
\includegraphics[scale=0.45]{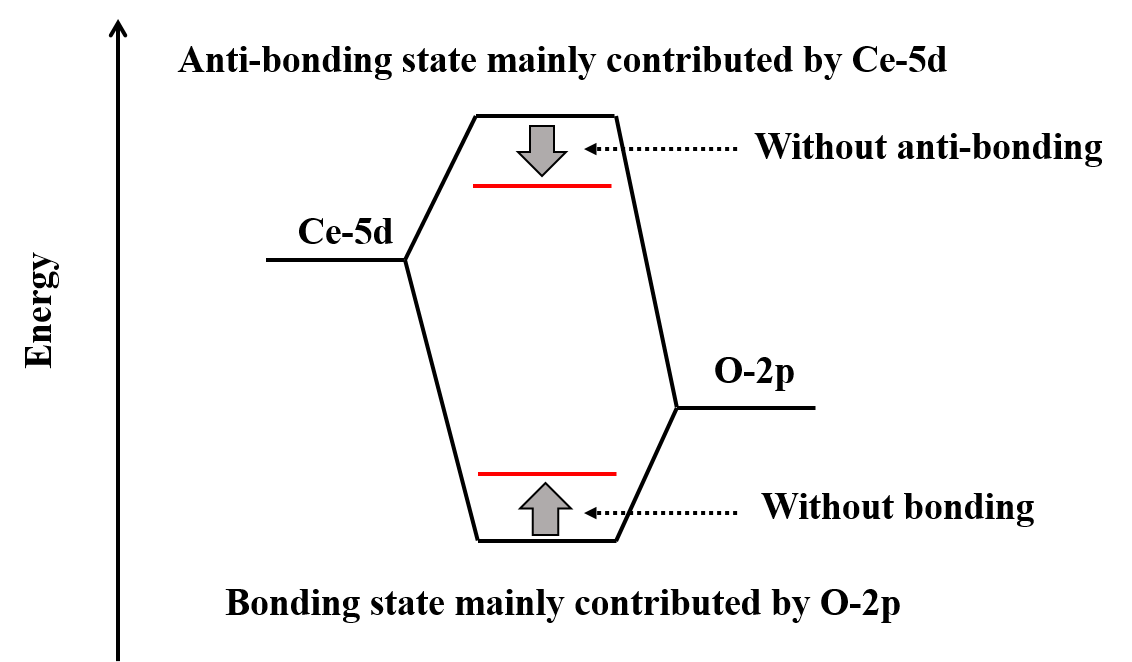}
\caption{Conceptual diagram to explain the red-shift induced by V$_O$: the Ce$_{5d}$ state is lowered by the absence of one oxygen atom}\label{figure_redshift}
\end{figure}

Alternatively, we try to explain the red-shift of LSO:Ce(1-a),V$_{O20}$ from the viewpoint of chemical bonding. The idea is shown in Fig.~\ref{figure_redshift}, indicating that the creation of neutral oxygen vacancy nearby the Ce$^{3+}$  ion will make the original bonding interaction between the O$_{2p}$ and Ce$_{5d}$ states disappear. The disappearance of such bonding interaction can yield an increase of energy of V$_O$ state, 
which is indeed observed in Fig.~\ref{figure_LSO-Ce} and Fig.~\ref{figure_LSO-Ce}, but also the decrease of energy of Ce$_{5d}$ state. Such idea had also been proposed to understand the red shift of BaMgAl$_{10}$O$_{17}$:Eu,V$_O$.\cite{BAM}

\begin{table}
\centering
\renewcommand\arraystretch{1.5}
\caption{Dorenbos model analysis of the [Xe]5d state of Ce$^{3+}$ ion in LSO, with and without V$_O$. $D$: Spectroscopic red-shift; GS: ground state; EX: excited state.}
\label{table-3}\begin{tabular}{l|c|c}\hline
Case& LSO:Ce1 & LSO:Ce(1-a),V$_{O20}$ \\ \hline
 $\chi_{av}$ & 1.44 & 1.44 \\ \hline
$\alpha_{sp}^N$ & 2.31 & 2.31 \\ \hline
$\varepsilon_{c}$, GS& 10260 cm$^{-1}$ & 8830 cm$^{-1}$ \\ \hline
$\varepsilon_{c}$, EX& 12008 cm$^{-1}$ & 9295 cm$^{-1}$ \\ \hline
 $\beta$& 1.22$\times$10$^9$ & 1.35$\times$10$^9$ \\ \hline
$\varepsilon_{cfs}$, GS& 19740 cm$^{-1}$ & 21968 cm$^{-1}$ \\ \hline
$\varepsilon_{cfs}$, EX& 20711cm$^{-1}$ & 22254 cm$^{-1}$ \\ \hline
$D$, GS& 18485 cm$^{-1}$ & 16093 cm$^{-1}$ \\ \hline
$D$, EX& 20638cm$^{-1}$ & 16678 cm$^{-1}$ \\ \hline
\end{tabular}
\end{table}

After the analysis of the effect of a neutral oxygen vacancy on Ce$^{3+}$ luminescence, we analyze the effect of charged oxygen vacancies V$_O^{2+}$. As mentioned before, we focus on the V$_O^{2+}$ at O$_{20}$ and O$_{22}$ sites. Again, the CDFT method allows one to compute the transition energy in the six combinations of LSO:Ce,V$_O^{2+}$.  Typical electronic band structures, for LSO:Ce(1-a),V$^{2+}_{O22}$, are shown in Fig.~\ref{figure_LSO-Ce}(c). Compared to the results of Fig.~\ref{figure_LSO-Ce}(b), the ground state band structure, A$_g$ case, indicates that V$_O^{2+}$ provides a new empty band below the conduction band minimum. Such empty band could act as electron trap, which might lead to afterglow and TL phenomena. Beside being a trap state, such empty band also can work as a recombination center. Indeed, through the analysis of charge density, the electronic band structure of  A$_g^*$ indicates that the nature of optical transition in LSO:Ce,V$_O^{2+}$ is the transition from Ce$_{4f}$ to the empty band from V$_O^{2+}$. Such optical transition is quite different from the inter-configuration $4f-5d$ transition, and suffers from a large configuration coordinate shift in the relaxation of excited state. The comparison of band structure results of A$_g$ and A$_e$, A$_g^*$ and A$_e^*$ clearly shows the distinct energy levels of Ce$_{4f}$. The calculated transition energy of Stokes shift of six combinations of LSO:Ce,V$_O^{2+}$  are listed in the Table~\ref{table-4}. An unusual large Stokes shift is observed in all the six cases of LSO:Ce,V$_O^{2+}$. The absorption and emission energies cannot match with experimental data of the lower energy emission peak, indicating such LSO:Ce,V$_O^{2+}$  cannot contribute the emission intensity for the lower energy band.
Thus, we hypothesize that is normal conditions, the Fermi level is such that none of the oxygen vacancies 
is charged.

\begin{table}[h]
\centering\renewcommand\arraystretch{1.5}
\caption{Transition energies and Stokes shift of Lu$_2$SiO$_5$:Ce,V$_O^{2+}$.}
\label{table-4}
\begin{tabular}{lccc} \hline
                                      & Abs                 & Em               & Stokes shift \\ \hline
Ce1-a,V$^{2+}_{O20}$ & 3.36 eV           & 1.98 eV         & 11049 cm-1   \\
Ce1-b,V$^{2+}_{O20}$ & 3.11 eV           & 1.75 eV         & 10969 cm-1   \\
Ce1-a,V$^{2+}_{O22}$ & 2.85 eV           & 1.14 eV         & 13792 cm-1   \\
Ce1-b,V$^{2+}_{O22}$& 2.80 eV            & 1.24 eV         & 12502 cm-1   \\
Ce2-a,V$^{2+}_{O22}$ & 2.79 eV           & 1.20 eV         & 12824 cm-1   \\
Ce2-b,V$^{2+}_{O22}$ & 3.19 eV           & 1.22 eV         & 15889 cm-1    \\ \hline
\end{tabular}
\end{table}

The above results show that the creation of oxygen vacancy (neutral or charged) close to the Ce$^{3+}$ ion  change its luminescence. Before the assignment of the origin of the low-energy emission band, we further perform a check on the possibility to have an immediate nonradiative recombination.\cite{Jia-2017,book1,book2} 
This analysis relies on $\Lambda$, the ratio between the Franck-Condon shift in the excited state and the absorption energy. With $\Lambda$ above 0.25, the semi-classical energetics is such 
that non-radiative recombination after excitation can happen immediately by 
reaching the $f-d$ crossing point just after the excitation, before energy 
is released in the form of phonons.  The results are listed in Table~\ref{table-5}, indicating that 
this mechanism is rolled out for the neutral vacancies: the worst case is $\Lambda=0.13$. 
The Ce$^{3+}$ ion at the substitutional Lu sites 
in combination with a neutral Ce-V$_O$ can act as a luminescent center according to this criterion.
By contrast, $\Lambda$ is larger than 0.25 for four of the charged vacancies cases.

\begin{figure}[h]
\includegraphics[scale=0.75]{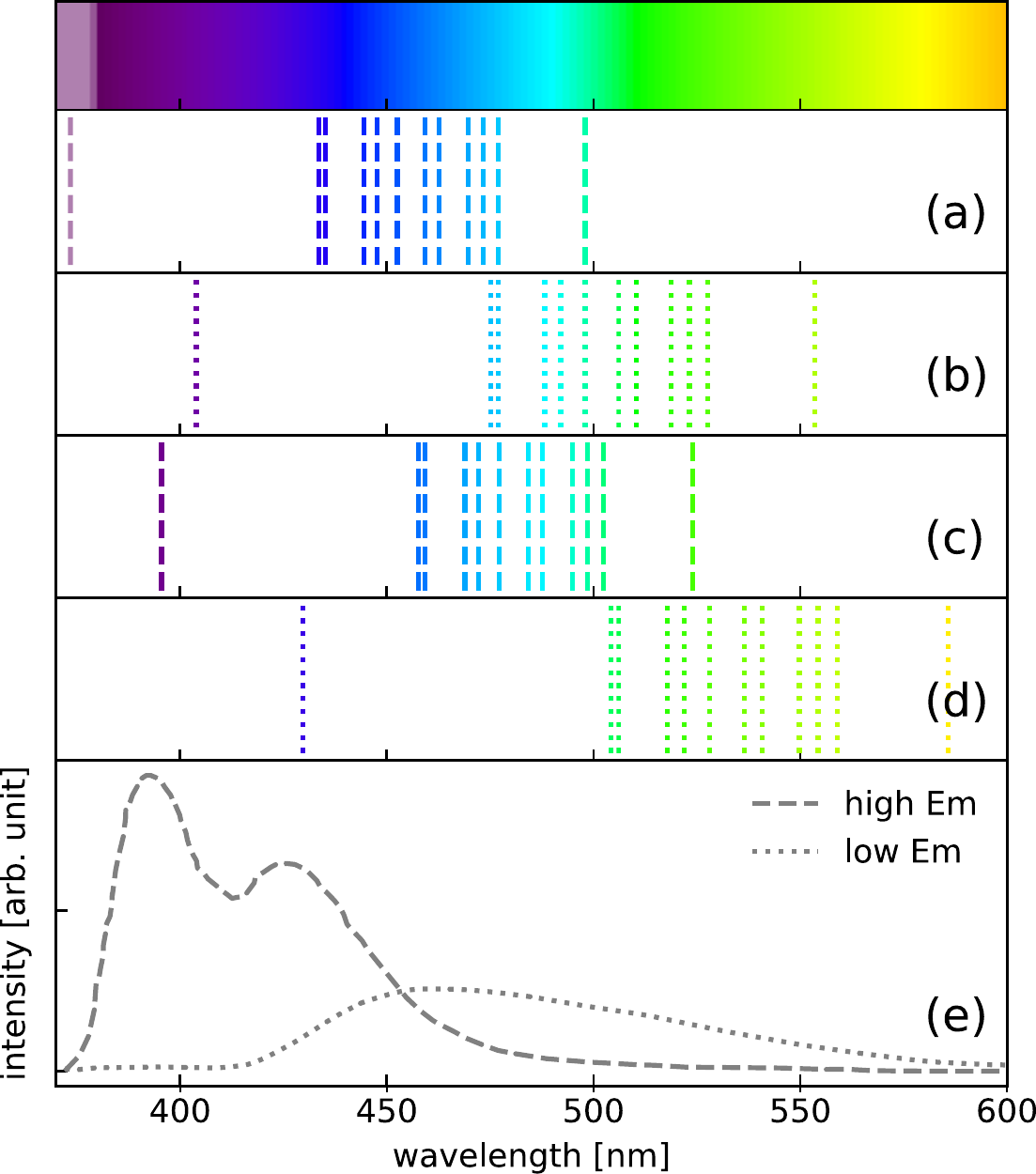}
\caption{Luminescent center assignment in LSO:Ce from first principles calculation. (a) $5d-4f (^2F_{5/2})$ transition energies from Table~\ref{table-1} and Table~\ref{table-2}; (b) $5d-4f (^2F_{7/2})$ transition energies. A simple constant energy  difference between $^2$F$_{7/2}$ and $^2$F$_{5/2}$ ground state levels,  of 2000~cm$^{-1}$, is considered; (c) corrected  $5d-4f (^2F_{5/2})$ transition energies; (d) corrected $5d-4f (^2F_{7/2})$ transition energies; (e) Experimental spectra at 11~K, from Ref. [\onlinecite{1993}].}
\label{figure_spectra}
\end{figure}

At this stage, we can assign the higher emission band to be from the Ce$^{3+}$ ion at the substitutional Lu1 site while the lower emission band in LSO:Ce ion is obtained from ten luminescent centers, the LSO:Ce2 and nine combinations of LSO:Ce,V$_O$ (be they from LSO:Ce1,V$_O$ or LSO:Ce2,V$_O$). The global picture from such assignment is shown in Fig.~\ref{figure_spectra}. First, the emission energies to the $^2$F$_{7/2}$ and $^2$F$_{5/2}$ states, with energy difference coming from the spin-orbit coupling, are
both considered, with a simple 2000~cm$^{-1}$ energy shift from the $^2$F$_{7/2}$ values
to obtain the $^2$F$_{5/2}$ values (cases (a) and (b) of Fig.~\ref{figure_spectra}). Then, these emission energies are further corrected by the previous overestimation of our methodology, which is found in our previous work about Ce$^{3+}$ doped phosphors\cite{Ce}, giving cases (c) and (d). The corrected emission energy of the LSO:Ce1 indeed nicely agrees with respect to the experimental value, both for $^2$F$_{7/2}$ and for $^2$F$_{5/2}$. The set of lines from LSO:Ce2 and nine combinations of LSO:Ce,V$_O$ cover a broad range, with a large overlap between the emission energies to the $^2$F$_{7/2}$ state and those to the $^2$F$_{5/2}$, so that 
the spin-orbit splitting effect cannot be seen.  Indeed, the range of emissions covered from the LSO:Ce2 and nine combinations of LSO:Ce,V$_O$ goes from 2.49 eV to 2.85 eV (difference of 0.36 eV), while the spin-orbit splitting is about 0.25 eV. This explains the characteristics of the emission spectra.

\begin{table}
\centering
\renewcommand\arraystretch{1.5}
\caption{Energy barrier E$_{fd}$(eV) and $\Lambda$ parameter for LSO:Ce with or without oxygen vacancy. $\Delta$C denotes the energy difference of the Franck-Condon shifts between excited and ground states, and ``\textit{x}" stands for the configuration coordinate for the $4f-5d$ crossover in the unit of the difference of excited (Q$_{e}$)and ground (Q$_g$) state coordinates.\cite{Jia-2017}  ``-" indicates that, in the parabolic approximation,  the $4f$ and $5d$ curves do not cross.}
\label{table-5}
\begin{tabular}{lcccccc}\hline
               & E$_{abs}$ & E$_{FC,e}$ & $\Delta$C &  \textit{1/x} & E$_{fd}$ & $\Lambda$ \\ \hline
Ce1                  & 3.804     & 0.221      & -0.046    & 0.183                         & 4.422    & 0.058     \\
Ce2                  & 3.619     & 0.316      & -0.133    & 0.298                         & 1.751    & 0.087     \\
Ce1-a,V$_{O20}$      & 3.069     & 0.21       & -0.008    & 0.154                         & 6.341    & 0.068     \\
Ce1-b,V$_{O20}$      & 3.45      & 0.348      & 0.106     & --                            & $\infty$ & 0.101     \\
Ce1-a,V$_{O21}$      & 3.333     & 0.237      & -0.068    & 0.231                         & 2.637    & 0.071     \\
Ce1-b,V$_{O21}$      & 3.287     & 0.277      & 0.041     & --                            & $\infty$ & 0.084     \\
Ce2,V$_{O21}$        & 3.274     & 0.427      & 0.074     & --                            & $\infty$ & 0.13      \\
Ce1,V$_{O11}$        & 3.352     & 0.221      & -0.125    & 0.27                          & 1.614    & 0.066     \\
Ce2,V$_{O11}$        & 3.105     & 0.233      & -0.218    & 0.311                         & 1.802    & 0.043     \\
Ce2-a,V$_{O02}$      & 2.985     & 0.22       & 0.06      & --                            & $\infty$ & 0.073     \\
Ce2-b,V$_{O02}$      & 3.173     & 0.245      & 0.057     & --                            & $\infty$ & 0.077     \\
Ce1-a,V$_{O20}^{2+}$ & 3.358     & 0.594      & -0.19     & 0.473                         & 0.735    & 0.177     \\
Ce1-b,V$_{O20}^{2+}$ & 3.107     & 0.605      & -0.15     & 0.488                         & 0.666    & 0.195     \\
Ce1-a,V$_{O22}^{2+}$ & 2.852     & 0.905      & 0.098     & 0.576                         & 0.492    & 0.317     \\
Ce1-b,V$_{O22}^{2+}$ & 2.797     & 0.8        & 0.038     & 0.547                         & 0.548    & 0.286     \\
Ce2-a,V$_{O22}^{2+}$ & 2.793     & 0.799      & 0.002     & 0.571                         & 0.451    & 0.286     \\
Ce2-b,V$_{O22}^{2+}$ & 3.192     & 1.102      & 0.239     & 0.556                         & 0.702    & 0.345    \\\hline

\end{tabular}
\end{table}

\subsection{Thermal quenching model}
\label{thermal}

Until now, we have examined the origin of the two emission bands in LSO:Ce based on the
value of the transition energy and a simple semi-classical criterion for being a luminescent center. 
Our analysis indicates that the higher emission band is from a single Ce$^{3+}$ ion at the substitutional Lu1 site, and the lower emission band is contributed by many luminescent centers, based on substitutional Ce$^{3+}$ ions (at Lu1 or Lu2 site) combined (for all centers except one) with an oxygen vacancy. In this section,
to complete the physical picture of the luminescence in LSO:Ce, we study the thermal quenching behavior of the two emission bands. The origin of strong thermal quenching performance of the low-energy emission band is discussed. 

At present, there are mainly two mechanisms to explain the thermal quenching properties of rare-earth doped luminescent materials, $4f-5d$ crossover and auto-ionization models. As mentioned in the Section~\ref{num_approach}, the energy barrier for the two models are E$_{fd}$ and E$_{dC}$, respectively. 

The information about E$_{fd}$ has been already obtained in view of its use for the
semi-classical luminescence criterion. All the computed quantities for the assigned luminescent centers, following the 1D-CCD framework, have been listed in Table~\ref{table-5}. It can be clearly noted that the calculated E$_{fd}$ are above 1.5~eV for all ten luminescent centers of the low-energy emission band, which cannot explain the strong thermal quenching experimental observation. The experimental thermal quenching temperature, $T_{0.5}$, of the lower energy band is below 80~K, indicating its thermal quenching barrier  must be smaller than 0.2~eV. As a result, we conclude that  $4f-5d$ crossover  should not be the major  mechanism for the strong thermal quenching of the low-energy emission band.  However, we note that the above analysis for the E$_{fd}$ just relies on a one-dimensional configuration coordinate diagram, which provides an upper bound of E$_{fd}$. Previous studies indeed have shown that a more complex picture, multi-phonon model with linear and quadratic coupling, and enlarged dimensionality, can give a much smaller energy barrier for the crossover than a one dimensional one.\cite{Bartram1986-1,Bartram1986-2} Therefore, the $4f-5d$ crossover model in not yet completely ruled out now, and more detailed investigations are needed.

The calculation of thermal barrier for the auto-ionization model, E$_{dC}$, is much more difficult than the one of E$_{fd}$. Indeed E$_{dC}$ relies on the difference between the energy of some localized level and the conduction band minimum, the energy of the latter being badly described within the DFT-PBE framework. A high-level computational methodology might allow us to get it, see e.g. the GW approximation used in our previous work.\cite{BSON-GW}

However, we will not aim to compute accurately the auto-ionization thermal barrier 
of the two emission bands in LSO:Ce, but instead, we will focus on the relative difference between the two emission bands.  
Following Eq.[\ref{E_dC}], $\varepsilon$(c) is the same for the eleven luminescent centers. Accordingly, the difference between E$_{dC}$ for the set of eleven luminescent centers
can be calculated from $\varepsilon$($Ce^{3+,*}/Ce^{4+}$). 
A smaller $\varepsilon$($Ce^{3+,*}/Ce^{4+}$) indicates a larger E$_{dC}$ within the auto-ionization model.

The calculated transition level for our eleven luminescent centers are listed in Table~\ref{table-6}. We first analyze the calculated results for the excited state geometry (Q$_e$). In this case, all values for the low-energy emission band of luminescent centers where a vacancy is involved are smaller than that of LSO:Ce1, the luminescent center for the high-energy emission band. However, we have discussed that the contribution from LSO:Ce2 should be small for the lower energy peak. Such a result contradicts the experimental thermal quenching of the lower emission band. By contrast, the result of LSO:Ce2 is slightly larger than that of LSO:Ce1. Thus, we can conclude that the auto-ionization model cannot be the dominant mechanism for strong thermal quenching of the low-energy emission band either. When the geometry is shifted from Q$_{e}$ to Q$_{dC}$ the result is not as clear cut. Such geometry relaxation makes the above trend inverse for several cases, such as LSO:Ce1-b,V$_{O20}$, which the $\varepsilon$($Ce^{3+,*}/Ce^{4+}$,Q$_{dC}$) being 0.15~eV larger than that of LSO:Ce1, with a smaller thermal barrier for the former than the latter. In principle the  ionized geometry should be the relevant one in a thermal equilibrium, thus the auto-ionization model cannot be completely discarded. However, it hardly explains the relative thermal quenching between the high-energy emission peak and the low-energy emission peak, since for the most of the cases, LSO:Ce,V$_O$ gives a larger E$_{dC}$ than that of LSO:Ce1.

\begin{table}[]
\centering\renewcommand\arraystretch{1.5}
\caption{Optical and thermodynamic transition level, $\varepsilon$($Ce^{3+,*}/Ce^{4+}$), in LSO:Ce. [Unit: eV]}
\label{table-6}
\begin{tabular}{lcc}\hline
Case                     & $\varepsilon(Ce^{3+,*}/Ce^{4+}$, $Q_e$) & $\varepsilon(Ce^{3+,*}/Ce^{4+}$, $Q_{dC}$)\\ \hline
Ce1                       & 7.55     &     8.04      \\
Ce2                       & 7.61     &     8.00      \\
Ce1-a,V$_{O20}$ & 6.88     &     7.47        \\
Ce1-b,V$_{O20}$ & 7.27     &     \textbf{8.19 }      \\
Ce1-a,V$_{O21}$ & 7.01     &     7.67         \\
Ce1-b,V$_{O21}$ & 6.98     &     7.67        \\
Ce2,V$_{O21}$    &  6.83    &     7.60         \\
Ce1,V$_{O11}$    & 6.94     &     8.00         \\
Ce2,V$_{O11}$    & 7.02     &     8.03         \\
Ce2-a,V$_{O02}$ & 7.10     &    7.66          \\
Ce2-b,V$_{O02}$ & 7.13     &    7.78          \\ \hline
\end{tabular}
\end{table}

\begin{figure}
\includegraphics[scale=0.40]{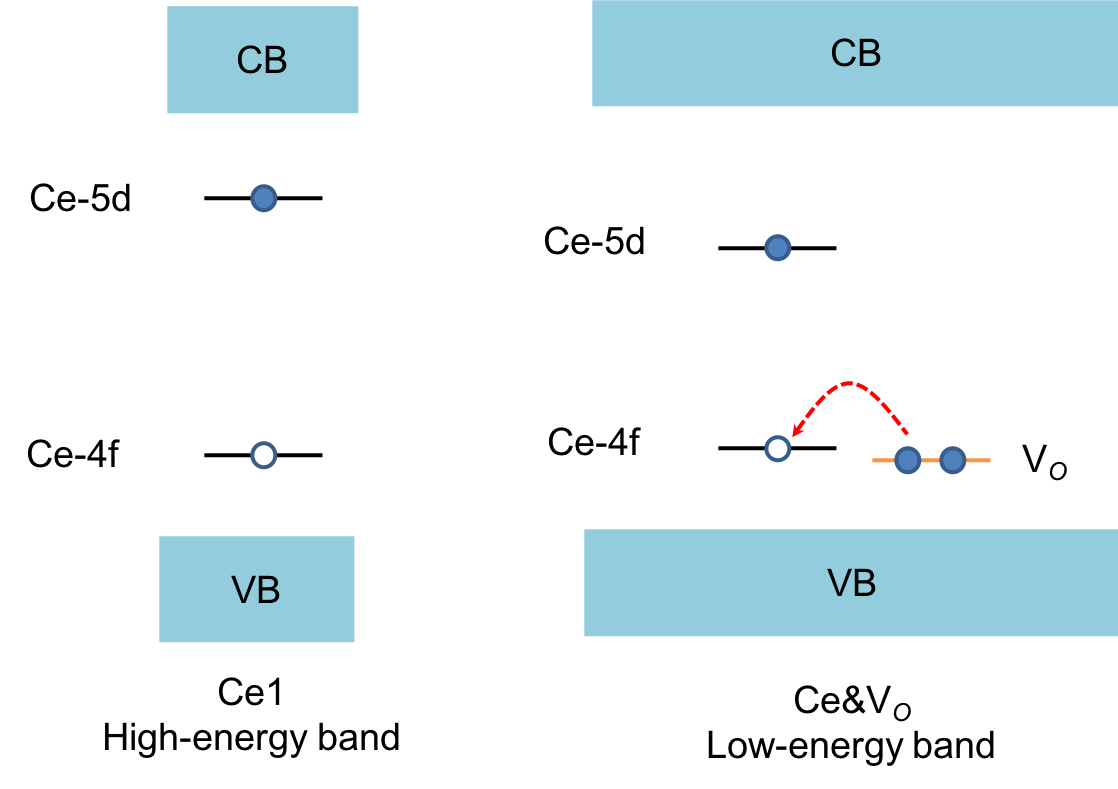}
\caption{Electron transport process for the thermal quenching behavior of LSO:Ce,V$_O$. The normal situation of LSO:Ce1 is shown for comparison.}\label{figure_hole}
\end{figure}

Based on the above analysis, the two well-known models for the thermal quenching seem inadequate
 to explain the large difference of thermal quenching between the two emission band in LSO:Ce. 
Thus, we put forward another thermal quenching mechanism for the low-energy emission band. 
The idea is based on the electronic band structures of LSO:Ce,V$_O$, and the equivalent role of charge carriers, free electron and hole, in the domain of semi-conductor physics. The results of A$_g^*$ and A$_e^*$ in Fig.~\ref{figure_LSO-Ce}(b) clearly show the overlap between the unoccupied Ce$_{4f}$ state and 
the occupied state from V$_O$.  Thus, it is reasonable to consider the auto-ionization process of the hole of Ce$_{4f}$ state to the occupied state of V$_O$.  The idea is depicted in the Fig.~\ref{figure_hole}.  Such hole auto-ionization process will result in the decrease of the recombination rate of $5d-4f$ transition for the luminescence. The difference of thermal quenching behaviors for the higher and low-energy emission band is due to the absence of V$_O$ state nearby the Ce1 site, that is the unique source of the higher band emission. Therefore, there is no such hole auto-ionization process in LSO:Ce1, resulting in a higher thermal stability of the high-energy emission band than the lower one, which is mainly contributed from LSO:Ce,V$_O$. 

To confirm our assumption, we performed additional calculations for the LSO:Ce$^{2+}$,V$_O^+$ case. In the calculation, the electron occupancies correspond to Fig.~\ref{figure_hole}, that is, both of Ce$_{4f}$ and Ce$_{5d}$ states are occupied by one single electron and an electron is removed from oxygen vacancy state. For the nine cases of LSO:Ce,V$_O$, the result indicates that there are three single occupied state inside the band gap, which are from V$_O^+$, Ce$_{4f}$ and Ce$_{5d}$ states, in increasing energy order. Table~\ref{table-7} list the total energies from the LSO:Ce$^{3+,*}$,V$_O$ (the one considered in the previous sections) and the LSO:Ce$^{2+}$,V$_O^+$
electronic states for the ground state (Q$_g$) and excited state (Q$_e$) geometries.  We find that the total energies of LSO:Ce$^{2+}$,V$_O^+$ is indeed lower than the one of LSO:Ce$^{3+.*}$,V$_O$, except just for the case of LSO:Ce1-a,V$_{O21}$, for which the total energy of LSO:Ce$^{2+}$,V$_O^+$ is slightly smaller than that of LSO:Ce$^{3+,*}$,V$_O$. These results justify the above-mentioned model we propose: After the excitation, the hole in the Ce$_{4f}$ states can be trapped by the neutral oxygen vacancy nearby. This trapping process will decrease the 5d-4f radiative-recombination while it increases the possibility of non-radiative processes. Such hole-transport deduced thermal quenching behaviour has been found in the photoluminescence of GaN semiconductor,\cite{PhysRevB.64.115205,GL-2014,reshchikov2014temperature} and recently was applied to the empirical analysis of the luminescence of rare earth ions in insulators.\cite{dorenbos5} Previous work indicated that the co-doping with divalent cation ions (Ca$^{2+}$) can significantly improve the scintillating performance of LSO:Ce:\cite{spurrier2008,2009} (i) shortening the scintillation decay time; and (ii) increasing light yield at the optimal concentration of divalent cation ions. We expect that such light yield enhancement in Ca$^{2+}$ codoped LSO:Ce single crystals can be attributed to the dissociation of spatially correlated Ce$^{3+}$ ions and oxygen vacancies.\cite{wu2018,ZHU20131,JIA2018372} The proposed hole auto-ionization process might be helpful to understand such effects.

\begin{table}
\centering\renewcommand\arraystretch{1.5}
\caption{Comparison of total energies of LSO:Ce$^{3+,*}$,V$_O$ and LSO:Ce$^{2+}$,V$_O^+$ for the ground state (Q$_g$) and excited state (Q$_e$) geometry.  The total energies are shown as the relative difference with respect to the E$_g^*$ value of LSO:Ce1-a,V$_{O20}$, -36432.21~eV. [Unit: eV]}
\label{table-7}
\begin{tabular}{lcccc} \hline

 & E$_g^*$   & Ce$^{2+}$-V$_O^+$,Q$_g$ & E$_e^*$    & Ce$^{2+}$-V$_O^+$,Q$_e$  \\ \hline
Ce1-a,V$_{O20}$ & 0.00          & -0.34 & -0.21 & -0.50 \\
Ce1-b,V$_{O20}$ & 0.43          & -0.02 & 0.24  & -0.21     \\
Ce1-a,V$_{O21}$ & 0.43          & 0.2   & 0.20   & \textbf{0.30 }      \\
Ce1-b,V$_{O21}$ & 0.59          & 0.58  & 0.36  & 0.33       \\
Ce2,V$_{O21}$   & 0.64           & 0.31  & 0.25  & 0.19        \\
Ce1,V$_{O11}$   & 0.40             & 0.21  & 0.18  & 0.13        \\
Ce2,V$_{O11}$   & 0.53           & 0.45  & 0.39  & 0.21        \\
Ce2-a,V$_{O02}$ & 0.38          & 0.32  & 0.22  & 0.21        \\
Ce2-b,V$_{O02}$ & 0.62          & 0.48  & 0.45  & 0.34       \\\hline

\end{tabular}
\end{table}

\section{Conclusion}
\label{conclusion}

In summary, we have studied from first principles the open issues in Lu$_2$SiO$_5$:Ce$^{3+}$ luminescence, dating from the nineties: the luminescent center identification and thermal quenching behaviour. To assign the luminescent center,  we simulate the $4f\rightarrow 5d$ neutral excitation of the Ce$^{3+}$ ions through a constrained density-functional theory method. The effect of oxygen vacancies on the luminescence of Ce$^{3+}$ ion is investigated here, and appears to be crucial. The calculation results indicate first that electronic and optical properties of the Ce$^{3+}$ ion depend noticeably on the adopted substitutional Lu site. Ce$^{3+}$ ions prefer to enter the Lu1 site, seven-coordinated by oxygen atoms. Further creation of oxygen vacancies close to these Ce$^{3+}$ ions can lead to a red shift of the spectra, irrespective of the substitutional Lu site involved. The reason for such red shift might be the disappearance of bonding interactions between O$_{2p}$ and Ce$_{5d}$ states. Based on the comparison between the experimental transition energy and the calculated values for LSO:Ce, the luminescent center for the high-energy emission band is due to the Ce$^{3+}$ ion substituting a Lu atom in the Lu1 site. The low-energy emission band originates from the contribution of ten Ce$^{3+}$ luminescent centers, one being LSO:Ce2 and nine being of LSO:Ce,V$_O$ type with different geometries (Ce1 as well as Ce2). 

For the thermal quenching behavior for the lower energy band, we analyze first  the role of two 
well-known mechanisms, $4f-5d$ crossover and electron auto-ionization. While we rely on the
existing one-dimensional configuration coordinate model for $4f-5d$ crossover, we focus
on differences of auto-ionization energies in the second case, and also investigate from first principles
the associated geometry relaxation effects, largely ignored in previous studies.
Our calculated results indicate that both  mechanisms have difficulties to explain its low thermal stability within our methodology. 
Beyond the two commonly considered mechanisms, we investigate from first principles the hole auto-ionization process, that might be the dominant mechanism for the strong thermal quenching of the low-energy emission band. 

The methodology and concepts of the present study can be applied to other luminescent materials in the areas of scintillators, LED phosphors, solid state lasers. 
We expect this to trigger reconsideration of the luminescence mechanisms in Lu$_2$SiO$_5$:Ce$^{3+}$
and other materials and be the basis for structure-property relationship analysis (emission color and thermal quenching behavior) of other luminescent materials.

\begin{acknowledgments}
We acknowledge the help of J.-M. Beuken and M. Giantomassi for computational matters. This work, done in the framework of ETSF (project number 551), has been supported by the Fonds de la Recherche Scientifique (FRS-FNRS Belgium) through a Charg\'e de recherches fellowship (Y. Jia) and the PdR Grant No. T.0238.13 - AIXPHO (X. Gonze). Computational resources have been provided by the supercomputing facilities of the Universit\'e catholique de Louvain (CISM/UCL) and the Consortium des Equipements de Calcul Intensif en F\'ed\'eration Wallonie Bruxelles (CECI) funded by the FRS-FNRS under Grant No. 2.5020.11.
\end{acknowledgments}

\bibliography{LSO} 

\end{document}